\documentclass[conference]{IEEEtran}
\usepackage{url}
\IEEEoverridecommandlockouts
\usepackage{cite}
\usepackage{amsmath,amssymb,amsfonts}
\usepackage{algorithmic}
\usepackage{graphicx}
\usepackage{float}   
\usepackage[compatibility=false]{caption}
\usepackage{subcaption}
\usepackage{textcomp}
\usepackage{xcolor}
\usepackage[version=4]{mhchem}  
\usepackage{stfloats}
\usepackage{hyperref}
\usepackage{eso-pic}

\def\BibTeX{{\rm B\kern-.05em{\sc i\kern-.025em b}\kern-.08em
    T\kern-.1667em\lower.7ex\hbox{E}\kern-.125emX}}
\begin{document}


\title{\vspace{-0.8cm}Can Optimal Dispatch Models Recreate Reality? \\ A Retrospective Analysis of Europe’s 2022 Energy Crisis Using PyPSA-Eur\\ \vspace{-0.40em}
}

\author{
  \IEEEauthorblockN{Lukas Karkossa\textsuperscript{1,*},
    Marco Saretta\textsuperscript{1,2},
    Frederik Erhard Gullach\textsuperscript{3,4},
    Marta Victoria\textsuperscript{1}}
  \vspace{0.05cm}
  \IEEEauthorblockA{
    \textit{
      \textsuperscript{1}Technical University of Denmark,
      \textsuperscript{2}Ramboll Denmark,
      \textsuperscript{3}Aarhus University,
      \textsuperscript{4}COWI
    }
  }
  \IEEEauthorblockA{
    \{lalka, mvipe\}@dtu.dk, mcsr@ramboll.com, fegu@mpe.au.dk, \textsuperscript{*}Corresponding author
  }
\vspace{-0.8cm}
}

\maketitle
\enlargethispage{-1.5\baselineskip}
\newlength{\copyrightwidth}
\setlength{\copyrightwidth}{0.97\textwidth} 

\AddToShipoutPictureBG*{%
  \AtPageLowerLeft{%
    \hspace*{\dimexpr(\paperwidth-\copyrightwidth)/2\relax}%
    \raisebox{1.7cm}{%
      \parbox{\copyrightwidth}{%
        \centering
        \fontsize{9}{9}\selectfont
        979-8-3195-3554-2/26/\$31.00~\copyright{}2026 IEEE Personal use of this material is permitted. Permission from IEEE must be obtained for all other uses, in any current or future media, including reprinting/republishing this material for advertising or promotional purposes, creating new collective works, for resale or redistribution to servers or lists, or reuse of \newline any copyrighted component of this work in other works. 
      }%
    }%
  }%
}
\begin{abstract}

Electricity prices result from the complex interplay of supply and demand, which depends on variable renewable energy production, fuel costs, \ce{CO2} price, and grid bottlenecks. 

Between 2020 and 2024, COVID-19 shifted demand and disrupted supply chains and operations in Europe, while Russia's invasion of Ukraine further constrained gas supply, causing exceptional volatility in gas prices. It remains unclear whether optimal dispatch models can reliably replicate historical hourly prices during crises, given the rapid fluctuations in fuel prices and operators' limited foresight.

In this work, we ask whether an optimal dispatch model, parametrised with historical data on demand, fuel, and \ce{CO2} prices, can reproduce the observed market outcomes during this period. We perform hourly hindcasts of electricity generation in 35 countries from 2020 to 2024 using PyPSA-Eur and compare the nodal marginal electricity prices with historical ENTSO-E market prices using the Symmetric Mean Absolute Percentage Error (SMAPE). The scenarios compare static vs dynamic assumptions on fuel and CO$_2$ prices, as well as perfect foresight vs a two-week rolling-horizon optimization.

Combining high‑resolution fuel and \ce{CO2} price time series with limited‑foresight substantially improves hindcast accuracy, yielding an average SMAPE of 20.76\% based on the daily load-weighted average price for the entire Europe. While improvements relative to the scenario with perfect foresight and static price inputs occur, they are most pronounced during periods of high fuel‑price volatility, when marginal‑cost swings propagate to electricity prices.  In 2022, the optimal generation mix in most countries shows substantially less natural gas and more coal than historically observed. Other discrepancies can be attributed to the model’s omission of real‑world policy interventions, other dispatch constraints, and generator outages.

\end{abstract}


\linespread{1}
\section{Introduction}
Natural gas prices in Europe increased significantly in 2021-2022 for several reasons. In addition to geopolitical tension, some of them are supply constraints and the attempt to fill the European gas storage before the winter in 2022 \cite{Gunter}. This led to an energy crisis, with electricity prices reaching unprecedented levels. Studies of the energy crisis have examined the spillover effects of natural gas prices on the electricity price by linking the spillover to the mix of generators in individual countries \cite{CHULIA2024113862, Bajo2025}. While providing insights into which countries and system designs are more vulnerable, these studies neglect the effect of dispatch strategies. To better understand price formation, it is essential to assess generator dispatch strategies, as electricity prices are determined by operational costs and the dispatch decisions that meet demand. Optimal dispatch models are used to simulate this dispatch by minimizing the total system cost. These models can give valuable insights into system operation and electricity prices under different scenarios. In addition to optimal dispatch, performance depends on parameters such as the quality of the input data, the spatial and temporal representation, and the optimisation foresight. 

Validation can be done through hindcasting, which enables the comparison of model outputs with historical values. PyPSA-Eur is an open-source model of the European power system, and can be used as an optimal dispatch model. When PyPSA-Eur was introduced by Hörsch et al. \cite{HORSCH2018207}, it was done together with a validation of the network topology, total lines length, and renewable expansion potentials.
Further validation of the model for 2018 was performed by Unnewehr et al. \cite{UNNEWEHR2022}, who tested the effects of spatial resolution and import/export constraints and found that the model can capture the main power system characteristics. This validation only includes annual generator dispatch. Validation is also part of the PyPSA documentation, where the optimal operation was compared to historical values for 2019 using monthly average fuel prices\cite{PyPSAEurValidation}. This exercise identified several limitations, including an overestimation of renewable generation capacity factors, the absence of planned nuclear power plant shutdowns in the model, and the misclassification of some run-of-river plants. All of this led to differences in the mix of technologies generating electricity throughout the year. These issues highlight challenges in accurately representing renewable and conventional generation in the model. 
The energy crisis of 2022 presents an opportunity to validate optimal dispatch models in times where rapidly changing fuel and electricity prices can trigger demand response and potentially cause actual dispatch to diverge further from purely cost-minimising outcomes than under normal conditions. Based on these insights, this study poses the following research question:

\textit{Can PyPSA-Eur successfully hindcast the European energy crisis during the period 2020 to 2024 by reproducing similar energy and price dynamics?}

This work presents four novelties: (i) validating PyPSA-Eur during an energy crisis with unprecedented market conditions, (ii) employing high-resolution time series for fuel prices, (iii) analysing generator response during an energy crisis, and (iv) comparing the impacts of assuming perfect and limited foresight. The following sections present the methodology, followed by results and discussion.

\IEEEpubidadjcol

\enlargethispage{-1.5\baselineskip}
\section{Methods}
\label{sec:methods}
This section describes the modelling approach, data sources, scenario definitions, and validation methods used in this study.
\subsection{Model framework and scope}

An hourly electricity dispatch optimization was performed individually for the years 2020 to 2024 for 35 European countries, represented with a spatial resolution of 39 nodes using PyPSA-Eur v2025.07.0 \cite{HORSCH2018207}. No other sectors were included, e.g. heating, transport, or industry. Fixed generation and transmission capacities were assumed. Cross-border electricity exchanges were endogenously determined using linearized power flows and the clustered transmission network in PyPSA-Eur.
Hourly electricity demand was sourced from the ENTSO-E Transparency Platform \cite{entsoeTransparencyPlatform} and from national transmission system operator data for countries outside ENTSO-E (e.g., the UK \cite{nesoHistoricDemand}, Ireland \cite{soniElectricitySystem}). These historical demand profiles were imposed exogenously at each node. The model is formulated as a linear dispatch optimization problem. A load shedding price of 5,000 EUR/MWh was assumed to ensure feasibility. The model does not include unit commitment constraints such as start-up costs, minimum up/down times, or thermal cycling limitations.
Historical fuel and CO$_2$ emissions price data were collected from different sources. Daily time series were used for both natural gas \cite{gas_prices} and CO$_2$ emission allowance prices \cite{tradingeconomicsCarbonPermits}. For coal \cite{investingRotterdamCoal} and oil \cite{oil_prices} prices, weekly resolution was used. Lignite prices are scaled from coal prices, based on the observed period between 2019-2022. The marginal cost of each generator is computed from fuel prices, using the lower heating value, the generator's efficiency and the CO$_2$ price. For the static scenario, defined later, prices assumed for natural gas, coal, lignite and oil are 11.6, 7.8, 7.9, 26.8 EUR/MWh$_{\text{th.}}$ respectively, and CO$_2$ price of 50 EUR/tCO$_2$. 

\subsection{Generation capacities and availability}
Both conventional generation capacities and hydropower (run-of-river, reservoir hydro, pumped hydro storage) are based on  the combined database maintained at powerplantmatching v.0.7.1 \cite{GOTZENS20191} for the year 2020 and were kept constant across all simulations. Thermal generators are assumed to be fully available; not accounting for historical power plant outages, while nuclear generation is limited by a country specific capacity factor per year.

Renewable generation capacities are based on historical per-country data from IRENA \cite{irena2025renewable} and vary by year to capture the rapid expansion of renewables, especially solar PV. Renewable time series were obtained using atlite v0.4.1  \cite{Hofmann2021} to create capacity factor time series with 3 different resource classes for all solar and wind technologies per country, as well as hydropower inflow time series. ERA5 reanalysis data \cite{c3s_era5_2023} were used to calculate hydro inflow and wind speeds. For solar radiation, we used SARAH3 \cite{sarah_v003}.

\subsection{Perfect foresight vs. rolling horizon dispatch}
Perfect foresight solves the entire year in a single optimization problem, providing the model with full knowledge of demand, renewable availability, and fuel prices. This allows long-term strategic use of storage, reservoir hydropower and interconnections, beyond what real-world operators can anticipate.
Rolling horizon dispatch, by contrast, divides the full single-year optimisation into a sequence of smaller optimisation problems, each covering a limited window of \textit{n} time steps. After each window is solved, the horizon advances, and the problem is re-optimised using updated information. This might reflect better real-world decision-making, where forecasts for prices, weather, and demand are only available for a limited period ahead and are subject to uncertainty.

\subsection{Scenario design}
To assess the impact of foresight assumptions and price variability, we evaluate three scenarios: 
\begin{itemize}
    \item perfect foresight-static
    \item perfect foresight-dynamic 
    \item rolling horizon-dynamic.
\end{itemize}
In all scenarios, each year (8760 hours) from 2020 to 2024 is optimised independently. The perfect foresight-static and perfect foresight-dynamic scenarios both solve each year as a single optimization problem. The key difference is that the static scenario uses fixed fuel and CO$_2$ prices, whereas the dynamic scenario uses high‑resolution historical time series of fuel and emission prices.
The rolling horizon-dynamic scenario also uses this information, but limits the model’s foresight to a two‑week (336 hours) optimisation window without temporal overlap, passing the final storage states between windows. This constrains the model’s ability to anticipate long-term price fluctuations.

\subsection{Benchmarking and validation}
The model outputs are benchmarked against historical electricity prices retrieved from the ENTSO-E transparency platform \cite{entsoeTransparencyPlatform}. The primary validation variable is the hourly electricity price, while generation by technology is analysed to interpret differences in price formation and dispatch behaviour.
To quantify model performance, the Mean Absolute Error (MAE), Root Mean Squared Error (RMSE), and Symmetric Mean Absolute Percentage Error (SMAPE) are computed between the modelled and historical price time series at hourly resolution, with daily and weekly aggregations. Modelled electricity prices are given by the nodal marginal prices (shadow prices) of the power balance constraint at each node.

The MAE measures the average deviation in EUR/MWh and is directly interpretable, although it remains sensitive to large absolute errors. The RMSE penalises large deviations more strongly due to the squaring of errors, making it useful for highlighting extreme price mismatches. The SMAPE normalises errors by the mean of observed and simulated values, enabling fair comparison across years with substantially different price levels.



\section{Results}

All price results from the three scenarios are evaluated against the ENTSO-E hourly market-clearing prices from 2020 to 2024. Results include a comparison with aggregated historical prices for Europe, a global error assessment across scenarios and years, and a country-level evaluation for Germany and Spain.

\subsection{Europe-aggregated price results}
\label{subsec:results_1}
\begin{figure}
  \centering
  \includegraphics[width=\linewidth]{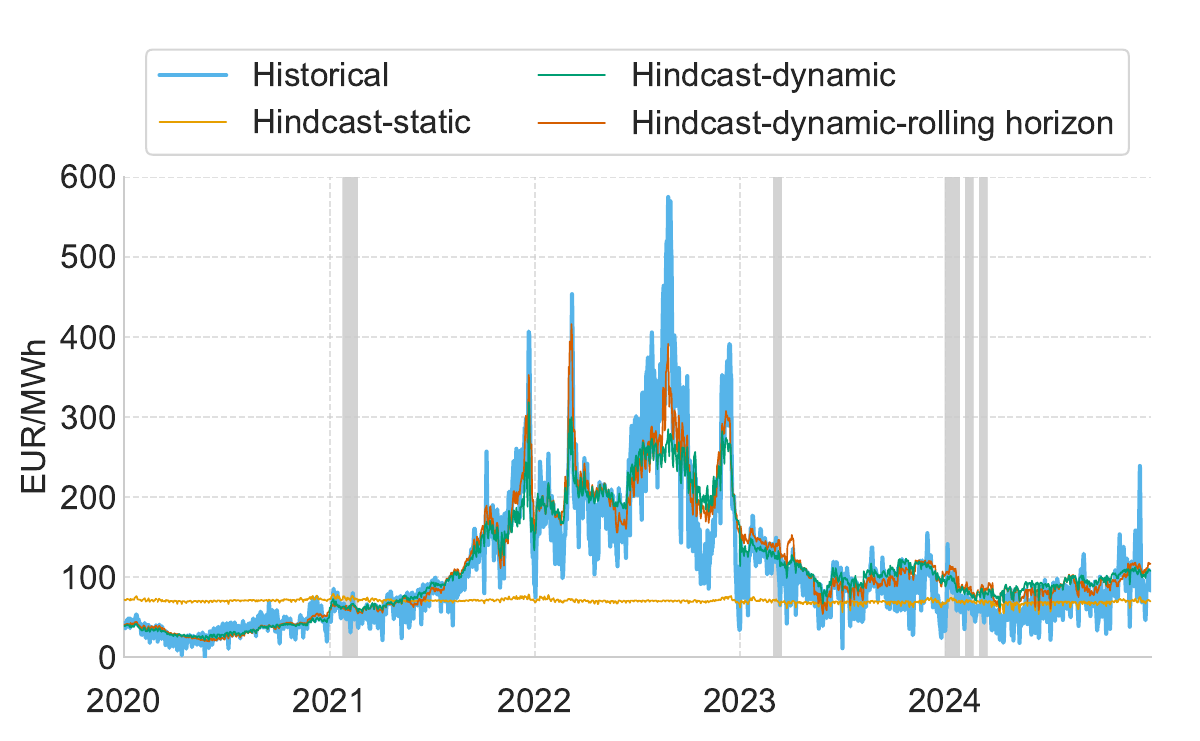}
  \caption{Europe-wide daily electricity price (load-weighted average across countries), 2020--2024.}
  \label{fig:Europe_price_filtered}
\end{figure}

Fig. \ref{fig:Europe_price_filtered} displays the European daily historical electricity prices for 2020–2024, compared with the simulation results. The European load-weighted average price was computed as the average of nodal prices across all countries, weighted by the respective load. Price results for the Hindcast-dynamic-rolling-horizon scenario were adjusted in the periods marked in grey due to load shedding.  During these intervals, the model triggered load shedding of up to 25\% of total demand in Norway due to the cyclic storage constraint, which enforces identical hydropower reservoir levels at the start and end of each optimisation window, hence causing infeasible dispatch in some intervals. To avoid distorting the aggregated average price, affected values were removed and replaced with the highest system price below a threshold of 1,000 EUR/MWh for countries exceeding this limit. This cleaned time series is also used for the following analysis.

From Fig. \ref{fig:Europe_price_filtered}, three main takeaways can be derived.
First, Dynamic scenarios capture market trends better than Static ones. Since the Static scenario is set up with fixed fuel and \ce{CO2} prices, it fails to reproduce intra-year market fluctuations. It overestimates electricity prices in 2020-2021, when COVID-induced low demand reduced fuel prices, and underestimates them in 2022-2024, when peak gas prices due to Russia's invasion of Ukraine created high historical electricity prices. The dynamic scenario, using daily/weekly historical fuel and emission prices, better reproduces historical trends, successfully capturing the general price trend from 2020 to 2024, including the large spikes in 2022.
Consistently, the SMAPE for the hourly European price drops from an average of 53.5 \% (Static) to 21.3 \% (dynamic) across 2020–2024. The improvement is most pronounced in 2022 (Static: 96.7 \% vs. dynamic: 22.6 \%).

Second, the Dynamic simulation is slightly outperformed by the Dynamic-rolling-horizon. Both Dynamic variants follow general trends, but the rolling horizon configuration matches the magnitude of historical spikes slightly better. While the hindcast dynamic optimises dispatch with perfect foresight, the rolling horizon version reacts short-sightedly, leading to less optimal, higher prices that align more closely with observed volatility. This is reflected in the SMAPE: the rolling horizon further reduces the multi-year average to 20.8\%. 

Third, the figure clearly shows daily price spikes and short-term price volatility are not captured well across all simulations, but both dynamic scenarios remain well below 100\% SMAPE based on hourly prices for all years, indicating that the modelled electricity prices are of the same order of magnitude as historically observed prices and that no scenario  systematically produces degenerate prices. 

\subsection{Global error matrix}

\begin{figure}[b]
  \centering
  \includegraphics[width=\linewidth]{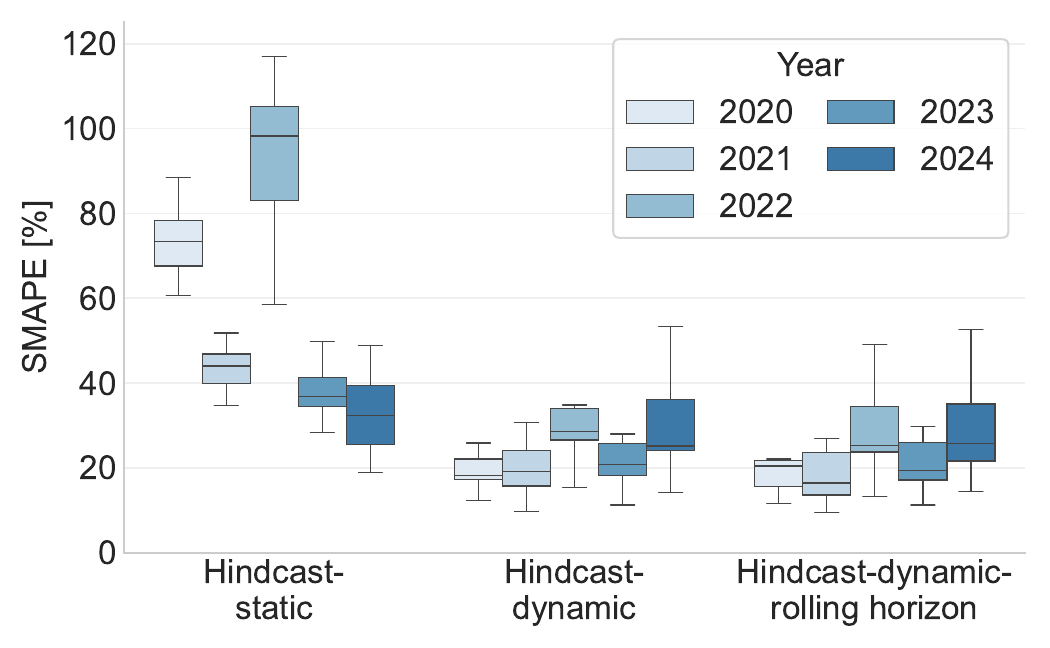}
  \caption{Boxplot of SMAPE across the years for all countries based on weekly prices. See Fig. \ref{fig:boxplot_smape_hourly} for SMAPE based on hourly prices.}
  \label{fig:error_smape_weekly}
\end{figure}
As a next step, nodal electricity prices were compared. Errors were calculated by comparing the simulated electricity prices from each model configuration with the corresponding historical prices on a weekly basis for all countries with available data. For the rolling horizon, the filtered prices were considered. In Fig. \ref{fig:error_smape_weekly}, the boxplots are organised into three clusters, each representing one scenario. Within each cluster, the five boxplots correspond to a year between 2020-2024, each boxplot containing the distribution of errors across the 24 countries used for validation in that year. Corresponding MAE and RMSE results are provided in the Supplementary Material \cite{github}.

The Static simulation performs worst, with median SMAPE reaching close to 100\% and upper whiskers exceeding 115\% during crisis periods, indicating that the Static setup is unable to represent historical data. This is inline with the results obtained for the European-aggregated time series described in section \ref{subsec:results_1}. 
In contrast, both dynamic configurations performed similarly and showed substantially lower errors on a weekly basis, with median SMAPEs of 20-26\%. The only notable difference is that the rolling horizon setup exhibits slightly wider whiskers, indicating greater variability, which might be due to load shedding. Overall, there is also a slight increase in accuracy applying the SMAPE on a weekly timescale, compared to hourly displayed in Fig. \ref{fig:boxplot_smape_hourly}.  

\subsection{Country case studies: Germany and Spain}


\begin{figure*}[t] 
\centering

\begin{subfigure}{\textwidth}
    \centering
    \includegraphics[width=0.9\linewidth]{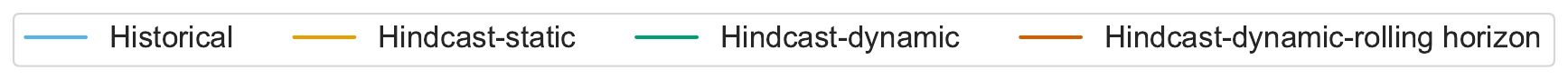}
\end{subfigure}

\vspace{0.1cm}

\begin{subfigure}{0.48\textwidth}
    \centering
    \includegraphics[width=\linewidth]{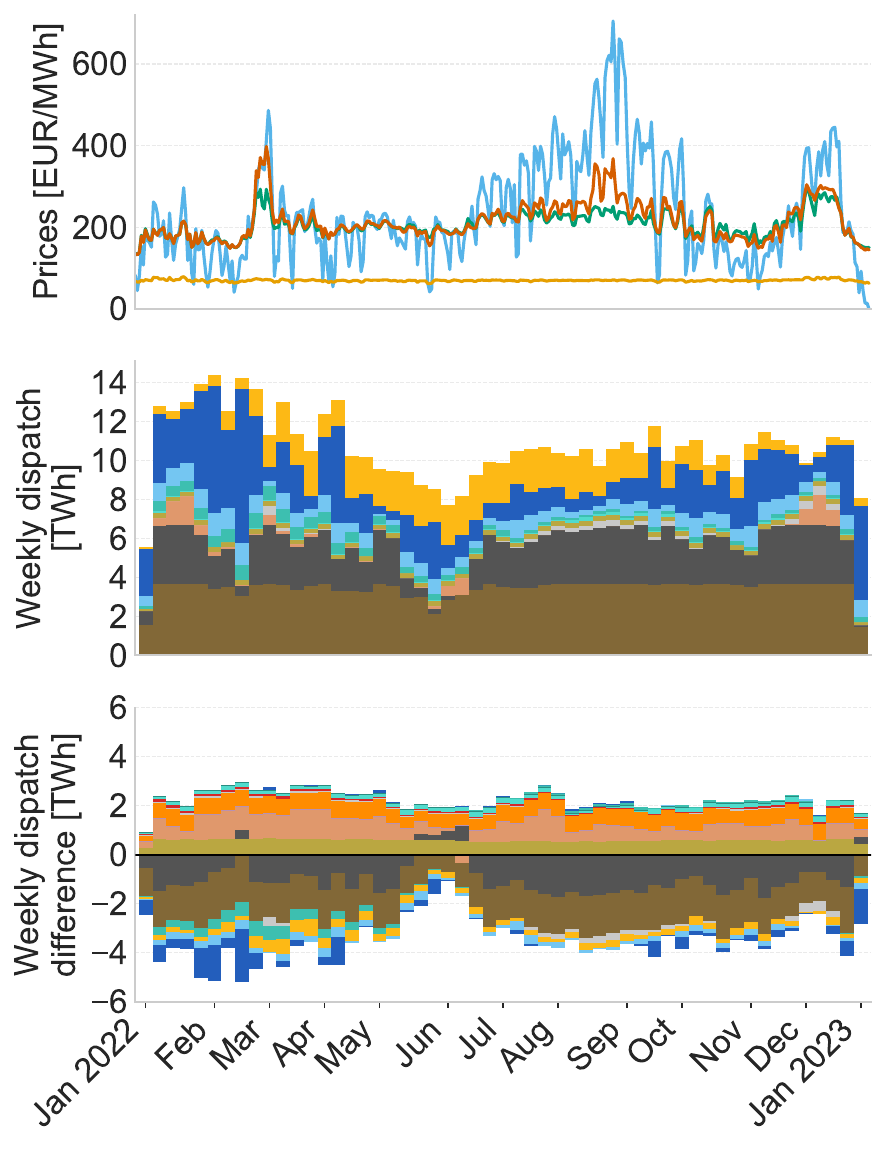}
    \caption{Germany.}
\end{subfigure}
\hfill
\begin{subfigure}{0.48\textwidth}
    \centering
    \includegraphics[width=\linewidth]{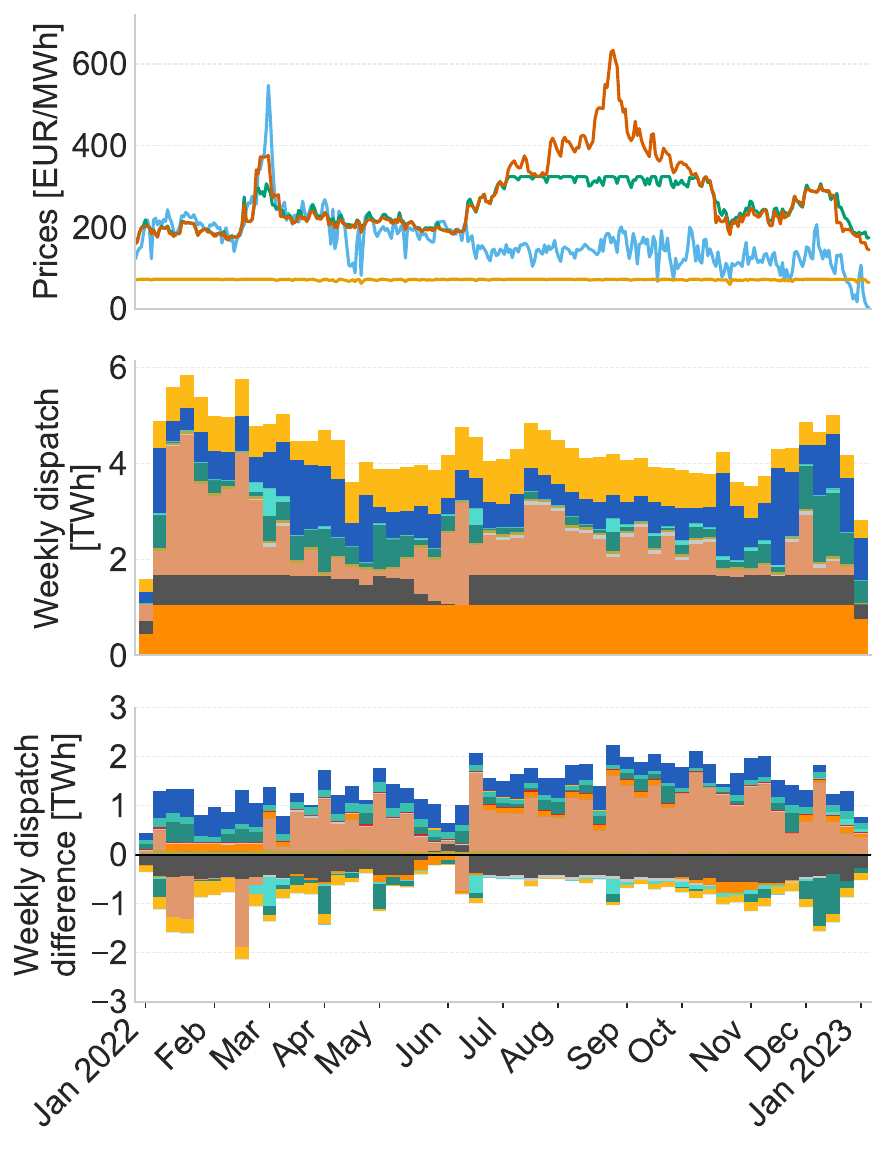}
    \caption{Spain.}
\end{subfigure}

\vspace{0.1cm}

\begin{subfigure}{\textwidth}
    \centering
    \includegraphics[width=0.9\linewidth]{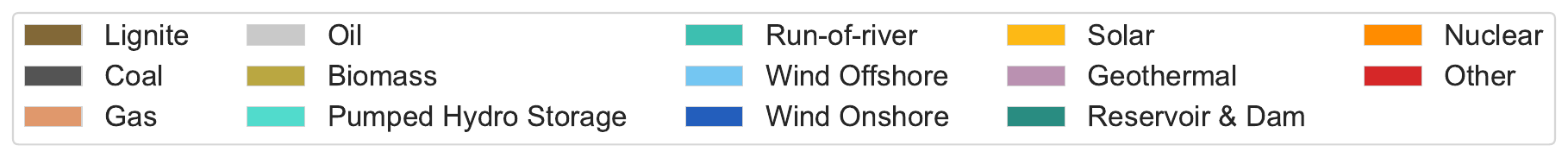}
\end{subfigure}

\caption{Country-level evaluation for Germany and Spain, year 2022. Top: Electricity price comparison. Middle: PyPSA-EUR generation by technology. Bottom: Difference between ENTSO-E and PyPSA (historical minus modelled) generation by technology. }
\label{fig:DE_ES_comparison_weekly}
\end{figure*}

    
    

    
    
    
Fig. \ref{fig:DE_ES_comparison_weekly} displays weekly prices and dispatch for 2022. The top panel compares simulated
and historical prices; the middle panel shows PyPSA-Eur generation by
technology for the rolling horizon scenario (best-performing); the bottom
panel shows the difference between ENTSO-E and PyPSA-Eur generation dispatch, where positive values indicate technologies generating more in historical data than in PyPSA-Eur.

\paragraph{Germany}
Comparing the different prices across the scenarios in the top panel of  Fig. \ref{fig:DE_ES_comparison_weekly} (a), one can see dynamic and dynamic rolling horizon scenarios broadly track the historical price trend in Germany, but consistently underestimate peak prices during the summer and autumn of 2022. In the dynamic rolling horizon scenario, gas-fired generation is largely avoided for most of the year. By contrast, ENTSO-E historical generation data indicates that gas was nearly used constantly throughout the year. Instead, the model systematically dispatches more coal and lignite. Wind and solar differences are small and are generally slightly overestimated, consistent with the difference in installed capacity (see Fig. \ref{fig:error_smape_weekly}). Nuclear and biomass generation are underestimated, as the model did not account for the 4 GW of nuclear capacity that was still in operation, according to historical statistics, and it also includes lower biomass capacity than historical.

\paragraph{Spain}
In Fig. \ref{fig:DE_ES_comparison_weekly} (b) top panel, a large deviation between historical and modelled prices can be seen. Essentially, the second half of the year seems to be decoupled. In the rolling horizon scenario, the limited two-week foresight yields more realistic reservoir and storage behaviour than the standard scenario. We observe a price spike of $\sim$ 600 EUR/MWh in August 2022 that did not occur in historical data. This is a direct consequence of the Iberian Exception \cite{robinson2023iberian}, a mechanism active from June 2022, under which Spain and Portugal capped the gas price used for electricity market clearing, decoupling Iberian prices from the broader European gas market. In the simulations, a uniform gas price across all nodes has been assumed, not accounting for this cap. As a result, gas-fired plant marginal costs in the model follow the gas price trajectory, driving the simulated spike that was not observed historically. This also explains most of the differences in dispatch: since gas was effectively cheaper in Spain due to the price cap, generators dispatched more gas than PyPSA-Eur, which instead reduced gas dispatch in response to higher assumed prices and used coal power plants to generate electricity. Other differences in dispatch patterns are partly due to reservoir hydropower following the cyclic storage constraint and therefore slightly differing from historical dispatch patterns, but mostly due to an underestimation of wind power capacity factors. This underestimation is not due to capacities, in fact installed capacity in PyPSA-Eur is slightly higher than reported by ENTSO-E, see Fig. \ref{fig:capacities_ES}. This might be caused by known ERA5 bias in the Iberian Peninsula, where reanalysis tends to underestimate wind speed \cite{MURCIA2022117794,UtraboCarazo2023}. 

\paragraph{Other countries}
Equivalent figures for Denmark, Norway, France, and Italy are provided in the supplementary material \cite{github}. Italy behaves similarly to Germany, dispatching less gas-fired generation throughout the year compared to historical levels. The difference is largely offset by coal-fired power plants.  In Norway, the cyclic storage constraint within the rolling-horizons leads to a complete break of the seasonal pattern of reservoir hydropower dispatch, which, combined with an underestimation of run-of-river capacity, leads to load shedding as discussed in Section \ref{sec:methods}. In France, during high-price periods, the model increases nuclear generation while reducing gas generation relative to historical data.

\section{Discussion and Limitations}


In the simplest scenario with perfect foresight and static fuel prices, the model does not reproduce the historic electricity price dynamics from 2020 to 2024. Allowing for time‑varying fuel prices enables the model to capture most trends and dynamics, but the largest electricity price peaks, especially in summer 2022, remain underestimated. None of the scenarios reproduced the observed intra-day price spikes. This gap could originate from the simplified marginal-cost pricing scheme: market prices are set by fuel costs, \ce{CO2} price, and variable O\&M costs, assuming uniformity across similar plants. In reality, unit-specific factors such as ramping constraints, efficiency differences, different fuel costs due to long-term agreements, startup costs, and outages drive additional variability.

The absence of such details can limit the model's ability to replicate real-time price volatility. Another explanation could be the differences in capacities assumed in the model and historical values. In the case of Germany, these differences should have contributed to a higher price, since biomass and nuclear capacities were underestimated, as seen in Fig. \ref{fig:capacities_DE}. The optimal generation mix in 2022 showed less gas and more coal than historical values in most countries; however, this trend is not sustained in the years before and after. Gas generation tends to be overestimated in Germany in those other years, whereas in Spain it fluctuates throughout the year. This raises the question of whether the market power exercised by different actors could have played a role.

Our analysis showed a small improvement in performance by reducing the full-year perfect foresight to a two-week rolling horizon. Two weeks of perfect foresight is still better than reality in a system where demand and supply should match in every instant, and both are linked to weather conditions.
Storage is implemented with a cyclic constraint (the state of charge at the start and end of each optimisation period is equal). In the rolling horizon scenario, this limits the ability of hydropower reservoir and pumped hydro storage plants to provide seasonal balancing.

In Norway, this led to large differences in hydropower operations relative to historical values and contributed to episodes of load shedding. Consequently, shortening the optimization horizon does not necessarily improve performance. The impact of this limitation is less severe in Spain, Fig. \ref{fig:DE_ES_comparison_weekly} (b). This is because the seasonality of hydropower operation is larger in Norway than in Spain, as can be seen in Fig. 2 in \cite{Gotske_Victoria_2021}
An alternative strategy to better reproduce seasonality under rolling horizon is to discard the cyclic storage constraint and impose a marginal storage value that represents the opportunity cost of the stored water \cite{Little1955StorageWater}, but this was not implemented here. 

Overall, our results are consistent with the validation in the PyPSA EUR documentation using data for 2019 \cite{PyPSAEurValidation}. Here, we extend the validation from dispatch to price formation and solve the model in rolling horizons over the more challenging 2020–2024 window. In both cases,  the dispatch of solar, onshore and offshore wind is slightly overestimated on a European scale. Nuclear generation is overestimated during summer (in 2020, 2022 and 2023) and underestimated during winter in all years. This is in line with the 2019 validation and could be due to the omission of planned nuclear outages \cite{PyPSAEurValidation}.

This study has several limitations. First, each country is represented by a single bidding zone, although in reality some countries consist of several bidding zones. This omits domestic constraints, which may translate to a distorted representation of market results at country level. Second, fuel prices were applied uniformly across all countries, neglecting national price differences arising from infrastructure, long-term supply contracts, and access to import routes. This also prevents representation of policy measures such as the “Iberian exception” which helps to explain deviations of our modelled prices from the historical prices in affected regions. Third, many thermal power plants operate as combined heat and power (CHP) units, meaning they primarily produce heat while also generating electricity. This co‑production can raise prices, and because our framework does not include sector coupling, we cannot capture this effect. Fourth, due to lacking high‑frequency price data for lignite, we applied a coal-indexed proxy using a historical scaling factor between coal and lignite.

\bibliographystyle{ieeetr} 
\newpage{}
\bibliography{references}

@misc{PyPSAEurValidation,
  author       = {{PyPSA-Eur Contributors}},
  title        = {Validation},
  howpublished = {\url{https://pypsa-eur.readthedocs.io/en/latest/validation.html}},
  year         = {2025},
  note         = {Accessed: 2026-01-29}
}

@article{Bajo2025,
author = {Bajo-Buenestado, Raul and Bento, Antonio and Kaffine, Daniel and Marmarelis, Zissis},
year = {2025},
month = {01},
pages = {170-181},
title = {Decarbonization and electricity price vulnerability},
volume = {8},
journal = {Nature Sustainability},
doi = {10.1038/s41893-024-01502-8}
}

@article{CHULIA2024113862,
title = {Vulnerability of European electricity markets: A quantile connectedness approach},
journal = {Energy Policy},
volume = {184},
pages = {113862},
year = {2024},
issn = {0301-4215},
doi = {https://doi.org/10.1016/j.enpol.2023.113862},
url = {https://www.sciencedirect.com/science/article/pii/S0301421523004470},
author = {Helena Chuliá and Tony Klein and Jorge A. {Muñoz Mendoza} and Jorge M. Uribe},
}

@article{Gunter,
author = {Güntner, Jochen and Reif, Magnus and Wolters, Maik},
title = {Sudden stop: Supply and demand shocks in the {German} natural gas market},
journal = {Journal of Applied Econometrics},
volume = {39},
number = {7},
pages = {1282-1300},
doi = {https://doi.org/10.1002/jae.3089},
url = {https://onlinelibrary.wiley.com/doi/abs/10.1002/jae.3089},
eprint = {https://onlinelibrary.wiley.com/doi/pdf/10.1002/jae.3089},
year = {2024}
}

@article{Hofmann2021, doi = {10.21105/joss.03294}, url = {https://doi.org/10.21105/joss.03294}, year = {2021}, publisher = {The Open Journal}, volume = {6}, number = {62}, pages = {3294}, author = {Hofmann, Fabian and Hampp, Johannes and Neumann, Fabian and Brown, Tom and Hörsch, Jonas}, title = {atlite: A Lightweight Python Package for Calculating Renewable Power Potentials and Time Series}, journal = {Journal of Open Source Software} }

@misc{entsoeTransparencyPlatform,
	author = {},
	title = {{T}ransparency {P}latform --- transparency.entsoe.eu},
	howpublished = {\url{https://transparency.entsoe.eu}},
	year = {},
	note = {[Accessed 02-10-2025]},
}

@misc{soniElectricitySystem,
	author = {},
	title = {{E}lectricity {S}ystem \&amp; {R}enewable {D}ata {R}eports | {T}he {G}rid | {S}{O}{N}{I} --- soni.ltd.uk},
	howpublished = {\url{https://www.soni.ltd.uk/grid/system-and-renewable-data-reports}},
	year = {},
	note = {[Accessed 13-10-2025]},
}

@misc{nesoHistoricDemand,
	author = {},
	title = {{H}istoric {D}emand {D}ata | {N}ational {E}nergy {S}ystem {O}perator --- neso.energy},
	howpublished = {\url{https://www.neso.energy/data-portal/historic-demand-data}},
	year = {},
	note = {[Accessed 14-10-2025]},
}

@misc{investingRotterdamCoal,
	author = {},
	title = {{R}otterdam {C}oal {F}utures {P}rice {T}oday - {I}nvesting.com --- investing.com},
	howpublished = {\url{https://www.investing.com/commodities/rotterdam-coal-futures}},
	year = {},
	note = {[Accessed 06-10-2026]},
}

@misc{oil_prices,
	author = {EU Comission},
	title = {{W}eekly {O}il {B}ulletin {P}rices {H}istory},
	howpublished = {\url{https://view.officeapps.live.com/op/view.aspx?src=https%3A%2F%2Fenergy.ec.europa.eu%2Fdocument%2Fdownload%2F906e60ca-8b6a-44e7-8589-652854d2fd3f_en%3Ffilename%3DWeekly_Oil_Bulletin_Prices_History_maticni_4web.xlsx&wdOrigin=BROWSELINK}},
	year = {},
	note = {[Accessed 06-10-2025]},
}

@misc{gas_prices,
	author = {Yahoo finance},
	title = {Natural Gas Apr 26 (NG=F)},
	howpublished = {\url{https://finance.yahoo.com/quote/NG%3DF/history/}},
	year = {},
	note = {[Accessed 06-10-2025]},
}

@misc{tradingeconomicsCarbonPermits,
	author = {},
	title = { 	{E}{U} {C}arbon {P}ermits - {P}rice - {C}hart - {H}istorical {D}ata - {N}ews  --- tradingeconomics.com},
	howpublished = {\url{https://tradingeconomics.com/commodity/carbon}},
	year = {},
	note = {[Accessed 06-10-2025]},
}

@book{irena2025renewable,
  author       = {{International Renewable Energy Agency (IRENA)}},
  title        = {Renewable Capacity Statistics 2025},
  year         = {2025},
  publisher    = {International Renewable Energy Agency},
  address      = {Abu Dhabi},
  isbn         = {978-92-9260-652-7}
}

@misc{sarah_v003,
  doi = {10.5676/EUM_SAF_CM/SARAH/V003},
  url = {https://wui.cmsaf.eu/safira/action/viewDoiDetails?acronym=SARAH_V003},
  author = {Pfeifroth, Uwe and Kothe, Steffen and Drücke, Jaqueline and Trentmann, Jörg and Schröder, Marc and Selbach, Nathalie and Hollmann, Rainer},
  title = {Surface Radiation Data Set - Heliosat (SARAH) - Edition 3},
  publisher = {Satellite Application Facility on Climate Monitoring (CM SAF)},
  year = {2023}
}

@misc{c3s_era5_2023,
  author       = {{Copernicus Climate Change Service (C3S)}},
  title        = {ERA5 hourly data on single levels from 1940 to present},
  year         = {2023},
  publisher    = {Copernicus Climate Change Service (C3S) Climate Data Store (CDS)},
  doi          = {10.24381/cds.adbb2d47},
  url          = {https://doi.org/10.24381/cds.adbb2d47},
  note         = {Accessed on 18-08-2025}
}

@article{UtraboCarazo2023,
  title = {A Spectral Analysis of Near‐Surface Mean Wind Speed and Gusts Over the {Iberian} Peninsula},
  volume = {50},
  ISSN = {1944-8007},
  url = {http://dx.doi.org/10.1029/2023GL103323},
  DOI = {10.1029/2023gl103323},
  number = {8},
  journal = {Geophysical Research Letters},
  publisher = {American Geophysical Union (AGU)},
  author = {Utrabo‐Carazo,  Eduardo and Azorin‐Molina,  Cesar and Aguilar,  Enric and Brunet,  Manola},
  year = {2023},
  month = apr 
}

@article{HORSCH2018207,
title = {{PyPSA-Eur}: An open optimisation model of the {European} transmission system},
journal = {Energy Strategy Reviews},
volume = {22},
pages = {207-215},
year = {2018},
issn = {2211-467X},
doi = {https://doi.org/10.1016/j.esr.2018.08.012},
url = {https://www.sciencedirect.com/science/article/pii/S2211467X18300804},
author = {Jonas Hörsch and Fabian Hofmann and David Schlachtberger and Tom Brown},
}

@article{GOTZENS20191,
title = {Performing energy modelling exercises in a transparent way - The issue of data quality in power plant databases},
journal = {Energy Strategy Reviews},
volume = {23},
pages = {1-12},
year = {2019},
issn = {2211-467X},
doi = {https://doi.org/10.1016/j.esr.2018.11.004},
url = {https://www.sciencedirect.com/science/article/pii/S2211467X18301056},
author = {Fabian Gotzens and Heidi Heinrichs and Jonas Hörsch and Fabian Hofmann}
}

@INPROCEEDINGS{UNNEWEHR2022,
  author={Unnewehr, Jan Frederick and Schäfer, Mirko and Weidlich, Anke},
  booktitle={2022 Open Source Modelling and Simulation of Energy Systems (OSMSES)}, 
  title={The value of network resolution – A validation study of the {European} energy system model PyPSA-Eur}, 
  year={2022},
  volume={},
  number={},
  pages={1-7},
  doi={10.1109/OSMSES54027.2022.9769123}}

@article{robinson2023iberian,
  title={The {Iberian} Exception: An overview of its effects over its first 100 days},
  author={Robinson, David and Arcos-Vargas, Angel and Tennican, Micheael and N{\'u}{\~n}ez, Fernando},
  journal={arXiv preprint arXiv:2309.02608},
  year={2023}
}

@article{Gotske_Victoria_2021,
  author = {Ebbe Kyhl Gøtske and Marta Victoria},
  title = {Future operation of hydropower in {Europe} under high renewable penetration and climate change},
  journal = {iScience},
  year = {2021},
  volume = {24},
  number = {9},
  pages = {102999},
  doi = {10.1016/j.isci.2021.102999},
  url = {https://doi.org/10.1016/j.isci.2021.102999}
}

@article{Little1955StorageWater,
  author  = {John D. C. Little},
  title   = {The Use of Storage Water in a Hydroelectric System},
  journal = {Journal of the Operations Research Society of America},
  year    = {1955},
  volume  = {3},
  number  = {2},
  pages   = {187--197},
  month   = {May},
  publisher = {INFORMS},
  url     = {https://www.jstor.org/stable/166669}
}

@misc{github,
	author = {},
	title = {{G}it{H}ub - marco-saretta/pypsa-network-analyzer: {R}epository to extract and visualise the main insights and plots from a {P}y{P}{S}{A} network file --- github.com},
	howpublished = {\url{https://github.com/marco-saretta/pypsa-network-analyzer}},
	year = {},
	note = {[Accessed 05-03-2026]},
}

@article{MURCIA2022117794,
title = {Validation of European-scale simulated wind speed and wind generation time series},
journal = {Applied Energy},
volume = {305},
pages = {117794},
year = {2022},
issn = {0306-2619},
doi = {https://doi.org/10.1016/j.apenergy.2021.117794},
url = {https://www.sciencedirect.com/science/article/pii/S0306261921011296},
author = {Juan Pablo Murcia and Matti Juhani Koivisto and Graziela Luzia and Bjarke T. Olsen and Andrea N. Hahmann and Poul Ejnar Sørensen and Magnus Als}
}
\enlargethispage{1\baselineskip}

\section*{Supplementary Material}

\renewcommand{\thefigure}{S\arabic{figure}}
\setcounter{figure}{0}

\begin{figure}[H] 
  \centering
  \includegraphics[width=\linewidth]{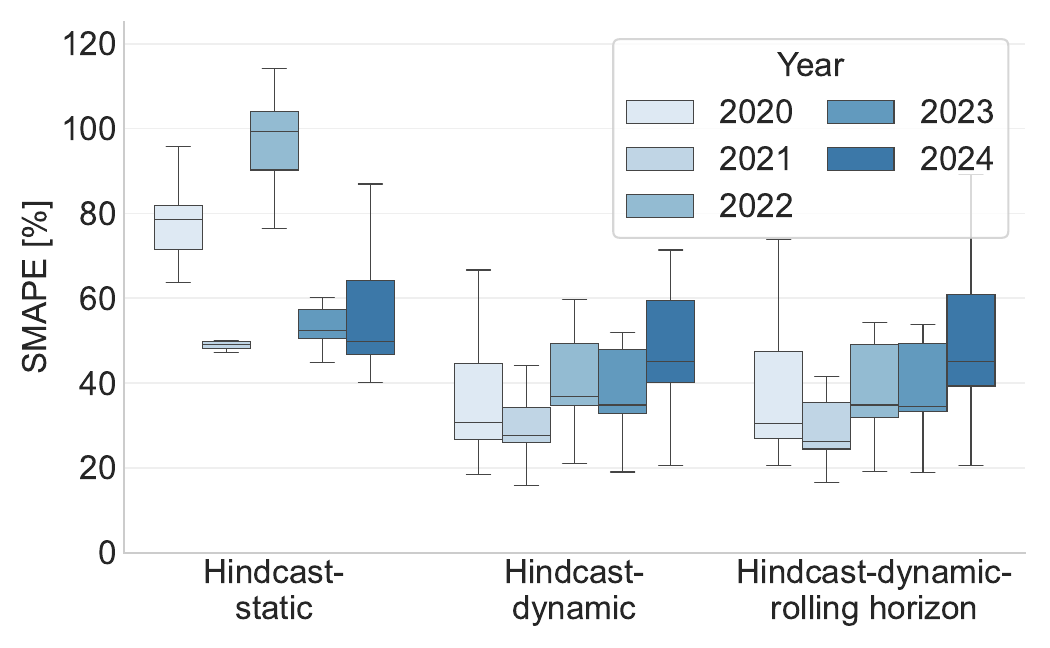}
  \caption{Boxplot of SMAPE across the years for all countries
based on hourly prices.}
  \label{fig:boxplot_smape_hourly}
\end{figure}

\begin{figure}[H]
  \centering
  \includegraphics[width=\linewidth]{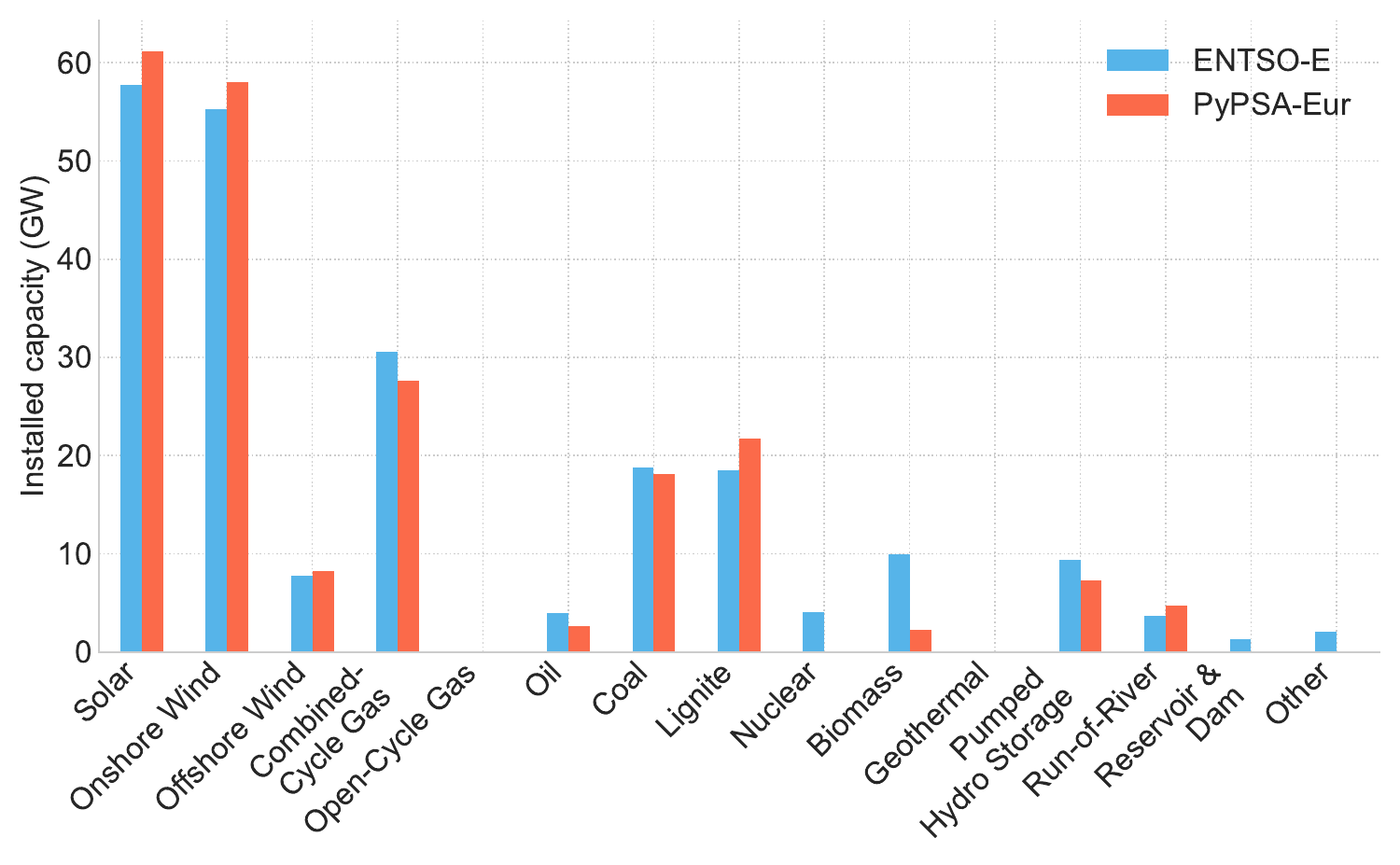}
  \caption{Installed generation capacities in Germany for 2022 as reported by ENTSO-E compared with the capacities used in the model.}
  \label{fig:capacities_DE}
\end{figure}

\begin{figure}[H]
  \centering
  \includegraphics[width=\linewidth]{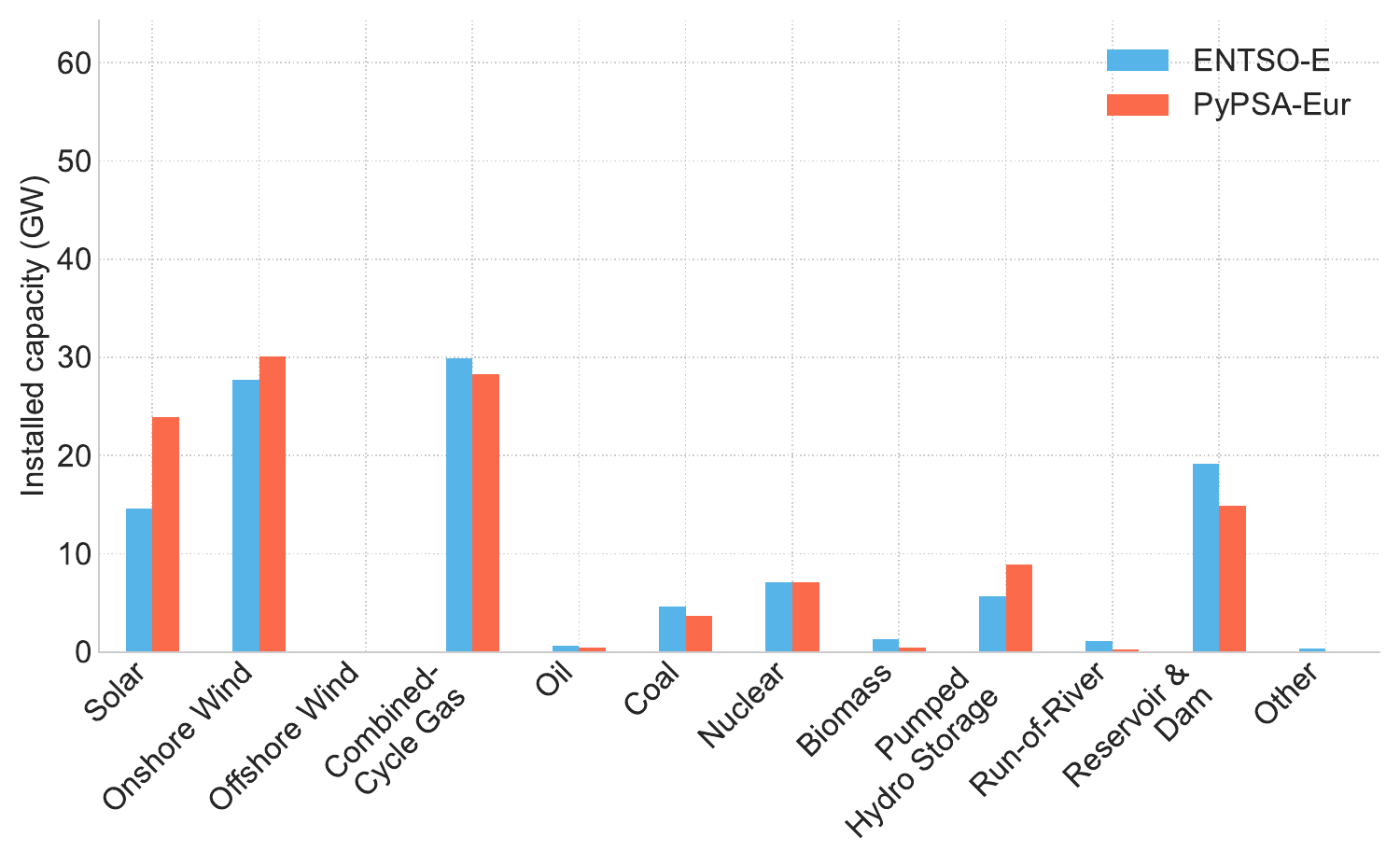}
  \caption{Installed generation capacities in Spain for 2022 as reported by ENTSO-E compared with the capacities used in the model.}
  \label{fig:capacities_ES}
\end{figure}

\end{document}